\magnification=\magstep1
\hoffset=0.5truecm
\voffset=0.5truecm
\hsize=16.5truecm 
\vsize=22.0truecm
\baselineskip=0.8 truecm
%\input today
%\baselineskip=0.6 truecm
%\baselineskip=14pt plus0.1pt minus0.1pt 
\parindent=25pt
\lineskip=4pt\lineskiplimit=0.1pt      
\parskip=0.1pt plus1pt

\let\st=\scriptstyle

%%%%%%%%%%%%%%%%%%%%%%%%%%%  FONTS

\font\twelvebf=cmbx12

\font\ninerm=cmr9

\font\ninebf=cmbx9

\font\sixrm=cmr6

%%%%%%%%%%%%%%%%%%%%%%%%% GRECO

 \let\b=\beta     \let\e=\varepsilon
  \let\h=\eta      \let\l=\lambda
   \let\n=\nu   \let\o=\omega      
  \let\s=\sigma \let\t=\tau   
  
\let\D=\Delta   \let\G=\Gamma  \let\L=\Lambda 
\let\O=\Omega

%%%%%%%%%%%%%%%%%%%%%%% CALLIGRAFICHE
%
\def\cA{{\cal A}}   
\def\cE{{\cal E}}  \def\cG{{\cal G}} 
   
\def\cM{{\cal M}}   \def\cP{{\cal P}}
 \def\cR{{\cal R}}  \def\cT{{\cal T}}

%%%%%%%%%%%%%%%%%%%%%%%%%%%%%%%%%% figure
%
\newdimen\xshift 
\newdimen\yshift
\newdimen\xwidth 
\def\eqfig#1#2#3#4#5#6{
  \par\xwidth=#1 \xshift=\hsize \advance\xshift 
  by-\xwidth \divide\xshift by 2
  \yshift=#2 \divide\yshift by 2
  \vbox{
  \line{\hglue\xshift \vbox to #2{
    \smallskip
    \vfil#3 
    \includegraphics{#4.ps}
    }
    \hfill\raise\yshift\hbox{#5}
  }
  \smallskip
  \centerline{#6}
  }
  \smallskip
}
%

%%%%%%%%%%%%%%%%%%%%%  Numerazione pagine
%
\def\data{\number\day/\ifcase\month\or gennaio \or febbraio \or marzo \or
aprile \or maggio \or giugno \or luglio \or agosto \or settembre
\or ottobre \or novembre \or dicembre \fi/\number\year}

%%\newcount\tempo
%%\tempo=\number\time\divide\tempo by 60}

\setbox200\hbox{$\scriptscriptstyle \data $}

\newcount\pgn 
\pgn=1
\def\foglio{\veroparagrafo:\number\pgn
\global\advance\pgn by 1}

%%%%%%%%%%%%%%%%% EQUAZIONI E TEOREMI CON NOMI SIMBOLICI

\global\newcount\numsec
\global\newcount\numfor
\global\newcount\numfig
\global\newcount\numtheo

\gdef\profonditastruttura{\dp\strutbox}

\def\senondefinito#1{\expandafter\ifx\csname#1\endcsname\relax}

\def\SIA #1,#2,#3 {\senondefinito{#1#2}%
   \expandafter\xdef\csname #1#2\endcsname{#3}\else
   \write16{???? ma #1,#2 e' gia' stato definito !!!!}\fi}

\def\etichetta(#1){(\veroparagrafo.\veraformula)
   \SIA e,#1,(\veroparagrafo.\veraformula)
   \global\advance\numfor by 1
   \write15{\string\FU (#1){\equ(#1)}}
   \write16{ EQ \equ(#1) == #1  }}

\def\FU(#1)#2{\SIA fu,#1,#2 }

\def\tetichetta(#1){{\veroparagrafo.\verotheo}%
   \SIA theo,#1,{\veroparagrafo.\verotheo}
   \global\advance\numtheo by 1%
   \write15{\string\FUth (#1){\thm[#1]}}%
   \write16{ TH \thm[#1] == #1  }}

\def\FUth(#1)#2{\SIA futh,#1,#2 }

\def\getichetta(#1){Fig. \verafigura
 \SIA e,#1,{\verafigura}
 \global\advance\numfig by 1
 \write15{\string\FU (#1){\equ(#1)}}
 \write16{ Fig. \equ(#1) ha simbolo  #1  }}

\newdimen\gwidth

\def\BOZZA{
 \def\alato(##1){
 {\vtop to \profonditastruttura{\baselineskip
 \profonditastruttura\vss
 \rlap{\kern-\hsize\kern-1.2truecm{$\scriptstyle##1$}}}}}
 \def\galato(##1){ \gwidth=\hsize \divide\gwidth by 2
 {\vtop to \profonditastruttura{\baselineskip
 \profonditastruttura\vss
 \rlap{\kern-\gwidth\kern-1.2truecm{$\scriptstyle##1$}}}}}
 \def\talato(##1){\rlap{\sixrm\kern -1.2truecm ##1}}
}

\def\alato(#1){}
\def\galato(#1){}
\def\talato(#1){}

\def\veroparagrafo{\ifnum\numsec<0 A\number-\numsec\else
   \number\numsec\fi}
\def\veraformula{\number\numfor}
\def\verotheo{\number\numtheo}
\def\verafigura{\number\numfig}

\def\Thm[#1]{\tetichetta(#1)}
\def\thf[#1]{\senondefinito{futh#1}$\clubsuit$[#1]\else
   \csname futh#1\endcsname\fi}
\def\thm[#1]{\senondefinito{theo#1}$\spadesuit$[#1]\else
   \csname theo#1\endcsname\fi}

\def\Eq(#1){\eqno{\etichetta(#1)\alato(#1)}}
\def\eq(#1){\etichetta(#1)\alato(#1)}
\def\eqv(#1){\senondefinito{fu#1}$\clubsuit$(#1)\else
   \csname fu#1\endcsname\fi}
\def\equ(#1){\senondefinito{e#1}$\spadesuit$(#1)\else
   \csname e#1\endcsname\fi}

% -------------------------------------------------------------------------
%
%  Numerazione verso il futuro ed eventuali paragrafi
%  precedenti non inseriti nel file da compilare
%
\def\include#1{
\openin13=#1.aux \ifeof13 \relax \else
\input #1.aux \closein13 \fi}
\openin14=\jobname.aux \ifeof14 \relax \else
\input \jobname.aux \closein14 \fi
\openout15=\jobname.aux

% -------------------------------------------------------------------------
%
% 

% -------------------------------------------------------------------------
%
\footline={\rlap{\hbox{\copy200}\ $\st[\number\pageno]$}\hss\tenrm
\foglio\hss}

% ---------------- fonti disponibili ---------------------------
%
\newcount\fnts
\fnts=0
\fnts=1 %-----comment if fonts msam, msbm, eufm are not available

%------------------------- Altre macro da chiamare ------------
%

\def\thsp{\thinspace}

\def\tthsp{\kern .083333 em}

\def\?{\mskip -10mu}

%------------------------ itemizing
%

\def\indbox#1{\hbox to \parindent{\hfil\ #1\hfil} }

\def\ref[#1]{[#1]}

\def\beginsubsection#1\par{\bigskip\leftline{\it #1}\nobreak\smallskip
	    \noindent}

\newfam\msafam
\newfam\msbfam
\newfam\eufmfam

% -------------------------------------------------- math macros --------
%
\ifnum\fnts=0% ------------Se non ci sono le fonti
  \def\bZ{ { {\rm Z} \mskip -6.6mu {\rm Z} }  }
  \def\bR{{\rm I\!R}}
  \def\bb{ \vrule height 6.7pt width 0.5pt depth 0pt }
  \def\bC{ { {\rm C} \mskip -8mu \bb \mskip 8mu } }
  \def\bE{{\rm I\!E}}
  \def\bP{{{\rm I\!P}}
  \def\mbox{
  \vbox{ \hrule width 6pt
     \hbox to 6pt{\vrule\vphantom{k} \hfil\vrule}
     \hrule width 6pt}
  }
  \def\QED{\ifhmode\unskip\nobreak\fi\quad
    \ifmmode\mbox\else$\mbox$\fi}
  \let\restriction=\lceil
\else% ------ o se ci sono
  \def\hexnumber#1{%
  \ifcase#1 0\or 1\or 2\or 3\or 4\or 5\or 6\or 7\or 8\or
  9\or A\or B\or C\or D\or E\or F\fi}
  \font\tenmsa=msam10
  \font\sevenmsa=msam7
  \font\fivemsa=msam5
  \textfont\msafam=\tenmsa
  \scriptfont\msafam=\sevenmsa
  \scriptscriptfont\msafam=\fivemsa        
  \edef\msafamhexnumber{\hexnumber\msafam}%
  \mathchardef\restriction"1\msafamhexnumber16
  \mathchardef\square"0\msafamhexnumber03
  \def\QED{\ifhmode\unskip\nobreak\fi\quad
    \ifmmode\square\else$\square$\fi}            
  \font\tenmsb=msbm10
  \font\sevenmsb=msbm7
  \font\fivemsb=msbm5
  \textfont\msbfam=\tenmsb
  \scriptfont\msbfam=\sevenmsb
  \scriptscriptfont\msbfam=\fivemsb
  \def\Bbb#1{\fam\msbfam\relax#1}    
  \font\teneufm=eufm10
  \font\seveneufm=eufm7
  \font\fiveeufm=eufm5
  \textfont\eufmfam=\teneufm
  \scriptfont\eufmfam=\seveneufm
  \scriptscriptfont\eufmfam=\fiveeufm
  \def\frak#1{{\fam\eufmfam\relax#1}}
  \let\goth\frak
  \def\bZ{{\Bbb Z}}
  \def\bR{{\Bbb R}}
  \def\bC{{\Bbb C}}
  \def\bE{{\Bbb E}}
  \def\bN{{\Bbb N}}
  \def\bP{{\Bbb P}}
  \def\bI{{\Bbb I}}
\fi
%
%-------------------------------------------------------------------
%
% ------- Per compatibilita'
%
\let\integer=\bZ
\let\real=\bR
\let\complex=\bC
\let\Ee=\bE
\let\Pp=\bP
\let\Dir=\cE
\let\Z=\integer
\let\uline=\underline
\def\Zp{{\integer_+}}
\def\ZpN{{\integer_+^N}}
\def\ZZ{{\integer^2}}
\def\ZZt{\integer^2_*}
\let\neper=e
\let\ii=i
\let\mmin=\wedge
\let\mmax=\vee
\def\identity{ {1 \mskip -5mu {\rm I}}  }
\def\ie{\hbox{\it i.e.\ }}
\let\id=\identity
\let\emp=\emptyset
\let\sset=\subset
\def\ssset{\subset\subset}
\let\setm=\setminus
\def\nep#1{ \neper^{#1}}
\let\uu=\underline
\def\ov#1{{1\over#1}}
\let\nea=\nearrow
\let\dnar=\downarrow
\let\imp=\Rightarrow
\let\de=\partial
\def\dep{\partial^+}
\def\deb{\bar\partial}
\def\tc{\thsp | \thsp}
\let\<=\langle
\let\>=\rangle
\def\xx{ {\{x\}} }
\def\xy{ { \{x,y\} } }
\def\pmu{\{-1,1\}}
\def\Pro{\noindent{\it Proof.}}
\def\sump{\mathop{{\sum}'}}
\def\tr{ \mathop{\rm tr}\nolimits }
\def\intt{ \mathop{\rm int}\nolimits }
\def\ext{ \mathop{\rm ext}\nolimits }
\def\Tr{ \mathop{\rm Tr}\nolimits }
\def\ad{ \mathop{\rm ad}\nolimits }
\def\Ad{ \mathop{\rm Ad}\nolimits }
\def\dim{ \mathop{\rm dim}\nolimits }
\def\weight{ \mathop{\rm weight}\nolimits }
\def\Orb{ \mathop{\rm Orb} }
\def\Var{ \mathop{\rm Var}\nolimits }
\def\Cov{ \mathop{\rm Cov}\nolimits }
\def\mean{ \mathop{\bf E}\nolimits }
\def\EE{ \mathop\Ee\nolimits }
\def\PP{ \mathop\Pp\nolimits }
\def\diam{\mathop{\rm diam}\nolimits}
\def\sign{\mathop{\rm sign}\nolimits}
\def\prob{\mathop{\rm Prob}\nolimits}
\def\gap{\mathop{\rm gap}\nolimits}
\def\tto#1{\buildrel #1 \over \longrightarrow}
\def\norm#1{ | #1 | }
\def\scalprod#1#2{ \thsp<#1, \thsp #2>\thsp }
\def\inte#1{\lfloor #1 \rfloor}
\def\ceil#1{\lceil #1 \rceil}
\def\intl{\int\limits}
\outer\def\nproclaim#1 [#2]#3. #4\par{\medbreak \noindent
   \talato(#2){\bf #1 \Thm[#2]#3.\enspace }%
   {\sl #4\par }\ifdim \lastskip <\medskipamount 
   \removelastskip \penalty 55\medskip \fi}
\def\thmm[#1]{#1}
\def\teo[#1]{#1}

%------------------------------ tilde
%
\def\sttilde#1{%
\dimen2=\fontdimen5\textfont0
\setbox0=\hbox{$\mathchar"7E$}
\setbox1=\hbox{$\scriptstyle #1$}
\dimen0=\wd0
\dimen1=\wd1
\advance\dimen1 by -\dimen0
\divide\dimen1 by 2
\vbox{\offinterlineskip%
   \moveright\dimen1 \box0 \kern - \dimen2\box1}
}
\def\ntilde#1{\mathchoice{\widetilde #1}{\widetilde #1}%
   {\sttilde #1}{\sttilde #1}}

%-------------------------------------------------------------------
%
%\sezioniseparate %----------- togliere quando si stampa tutto insieme
%\BOZZA %   
%\let\g=\o %      %------------------ per il Mac 

%\input $HOME/tex/for_enzo.tex 
%\BOZZA
%Macro
%rect 'rettangolo'
\def\rect#1#2{{\vcenter{\vbox{\hrule height.3pt
	    \hbox{\vrule width.3pt height#2truecm \kern#1truecm
	    \vrule width.3pt}
	    \hrule height.3pt}}}}
\def\square{\rect{0.15}{0.15}}
\def\plusone{+\underline{1}}
\def\minusone{-\underline{1}}
%Macro per l'inserimento della figura
\newdimen\xshift \newdimen\xwidth \newdimen\yshift \newdimen\ywidth
\def\ins#1#2#3{\vbox to0pt{\kern-#2 \hbox{\kern#1 #3}\vss}\nointerlineskip}
\def\eqfig#1#2#3#4#5{
\par\xwidth=#1 \xshift=\hsize \advance\xshift 
by-\xwidth \divide\xshift by 2
\yshift=#2 \divide\yshift by 2
\line{\hglue\xshift \vbox to #2{\vfil
#3 \includegraphics{#4.ps}}\hfill\raise\yshift\hbox{#5}}}

%Esempi utili
% Footnote: per esempio\footnote{$^1$}{testo della nota}

%%%%%%%%%%%%%%%%%%%%%%%Inizio lavoro - Beginning
%Sezione 0
\expandafter\ifx\csname sezioniseparate\endcsname\relax%
\input macro \fi
\numsec=0
\numfor=1\numtheo=1\pgn=1
%\hskip4truein \today
\par
\par
\noindent
\centerline {\twelvebf Metastability in the two-dimensional Ising model}
\centerline {\twelvebf with free boundary conditions}
\vskip 2.5 truecm
\centerline {Emilio N. M. Cirillo\footnote{$^\dagger$}{\noindent\tenrm
Permanent address: Dipartimento di Fisica dell'Universit\`a di Bari and
INFN, Sezione di Bari,
v. Amendola 173, I-70126 Bari, Italy}, Joel L. Lebowitz}
\centerline {\it Department of Mathematics and Physics, Rutgers University}
\par\noindent
\centerline {\it New Brunswick, NJ 08903, USA}
\par\noindent
\centerline {\rm E--mail: ecirillo@math.rutgers.edu, lebowitz@math.rutgers.edu}
\vskip 2.0 truecm
\centerline {\bf Abstract}
\vskip 0.5 truecm
\centerline{
\vbox{
\hsize=15truecm
\baselineskip=0.4cm
We investigate metastability in the two dimensional Ising model
in a square
with free boundary conditions at low temperatures. 
Starting with all spins down in a small positive magnetic
field, we show that the exit from this metastable phase occurs
via the nucleation of a critical droplet in one
of the four corners of the system. We compute the lifetime of the
metastable phase 
analytically in the limit $T\to 0$, $h\to 0$ and 
via Monte Carlo simulations 
at fixed values of $T$ and $h$ and find good agreement. 
This system models the effects of boundary domains 
in magnetic storage systems exiting
from a metastable phase when a small external field is applied.
}} 
\vfill\eject

\expandafter\ifx\csname sezioniseparate\endcsname\relax%
\input macro \fi
\numsec=1                
\numfor=1\numtheo=1\pgn=1
\vskip 1 truecm
\par\noindent
{\bf 1. Introduction.}
\par
Metastability is observed in
many different systems close to a first order phase transition. It is  
a dynamical phenomenon (see, for instance, [PL])
not included in the Gibbsian formalism, which is so successful for 
the description of stable equilibrium states [LR,I].
The development of a full theory of metastability is desirable for its
intrinsic as well as for
its experimental and technological interest and it also poses 
challenging mathematical problems (see [CGOV,OS1,OS2]).
\par
The metastable behavior of the nearest neighbor two dimensional Ising model  
for large finite volumes and small magnetic fields was analyzed in [NS1,NS2] 
in the zero temperature limit  
in the framework of the ``pathwise approach" introduced
in [CGOV]. In [S1] R. Schonmann, using arguments based on reversibility,
described in detail the typical escape paths from the metastable to the
stable regime. 
Other regimes, very interesting from the physical point of view and
mathematically much more complicated (finite temperature, infinite lattice and
zero magnetic field), are considered in [S2] and [SS].
The finite temperature case
has also been widely studied by Monte Carlo methods, for
instance in [B,BM,BS,TM1]; a complete and clear description of
these numerical results can be found in [RTMS]. This case has
also been studied by means of transfer-matrix and constrained-transfer-matrix
methods in [PS1,PS2,GRN].
\par
In the same asymptotic regime as in [NS1], different Ising-like 
hamiltonians have been considered in [KO1,KO2,NO] and 
the three dimensional nearest neighbor Ising model has been studied in [BC].
In [CO] the Blume-Capel model has been studied (see also [FGRN] for a 
study of a version of this model with weak long range interactions) and in [OS1,OS2] the problem of 
metastability has been investigated in a more general case.
\par
All the above works have been carried out for systems with 
periodic boundary conditions; in [RKLRN] Ising model has been studied in the
case of semiperiodic boundary conditions. 
In this note we study the
case of a finite lattice with free boundary conditions 
at low temperatures and small magnetic fields using both
rigorous analysis and Monte Carlo simulations.
\par
The case of free boundary conditions is of some technological interest:
during the recording process on magnetic tapes different parts of 
the magnetized material consisting of fine magnetic particles are exposed
to different magnetic fields, resulting in 
domains with different orientation of the magnetization. In order to be used
as storage devices, these materials must be able to retain their magnetization
for long periods in weak arbitrarily oriented magnetic fields.  
The study of the escape from a metastable phase in a periodic system
neglects the effects of boundary domains. Such effects are modeled here by 
considering a lattice with free boundary conditions. 
\par
We find that although the main features of the nucleation of the stable 
phase are
not changed, some interesting new aspects arise: in particular one
can say a priori where the nucleation of the
stable phase will start, that is where the critical droplet will show up.
\par
The model and results are described in Section 2 and analyzed 
in Section 3 and 4.
Section 5 is devoted to some brief conclusions.

\expandafter\ifx\csname sezioniseparate\endcsname\relax%
\input macro \fi
\numsec=2                
\numfor=1\numtheo=1\pgn=1
\vskip 1 truecm
\par\noindent
{\bf 2. The model and the results.}
\par
Let us consider a two dimensional Ising model defined on a finite
square $\L=\{1,...,M\}^2\subset\ZZ$ with free boundary conditions.
The space of
configurations is denoted by $\O=\{-1,+1\}^{\L}$ and to each
configuration $\s\in\O$ is associated the energy
$$
H(\s)=-{J\over 2}\sum_{<x,y>}\s(x)\s(y) -{h\over 2}\sum_{x\in\L}\s(x)
\;\;\;\;\;\;\;\;\;\s(x)=\pm 1\;\; ,
\Eq(hamiltonian)
$$
where the first sum runs over all the pairs of nearest neighbors in $\L$ 
and $J,h >0$.
The equilibrium states are described by the Gibbs measure 
$$
\n_{\L,\b}(\s)={e^{-\b H(\s)}\over
\sum_{\h\in\O} e^{-\b H(\h)}}\;\;\;\;\;\;\;\;\;\;\;\forall\s\in\O\;\; ,
\Eq(partition)
$$
where $\b$ is the inverse temperature.
\par
The time evolution of the model is given by the Metropolis algorithm: 
given a configuration $\s$ at time $t$, we pick a site $x$ at random and then
change $\s(x)$ to $-\s(x)$ with probability 1 if $\D H$ is $\le 0$
and $\exp(-\beta\;\D H)$ if $\D H > 0$. 
%as follows:
%given a configuration $\s_0$ at $t=0$ the configuration at time
%$t=1,2,...$ is obtained by using the Metropolis transition probability  
%$$
%p(\xi,\h)=P(\s_t=\h|\s_{t-1}=\xi)\;\;\;\;\; 
%\forall\xi,\h\in\O\;\; \forall t\ge 1
%$$
%with
%$$
%&p(\xi,\h)=\left\{\matrix{
%&{1\over |\L|}
%e^{-\b [\D H^x(\xi)]^+}&{\rm if}\;\exists x\in\L\; {\rm
%such\; that}\; \h=\xi^x\cr
%&0 &{\rm otherwise}\cr}\right.
%\Eq(transition)
%$$
%where
%$$
%\D H^x(\xi)=H(\xi^x)-H(\xi)\;\;\;\;\forall\xi\in\O\;\; 
%$$
%and $\xi^x$ is the configuration in which the spin at $x$ has been flipped:
%$$
%\xi^y(x)=\left\{\eqalign{
%-&\xi(x)\;\;{\rm if}\; x=y\cr
%&\xi(x)\;\;{\rm otherwise}\cr}\right.\;\; .
%\Eq(spin-flip)
%$$
\par
It is easy to show that this dynamics is reversible with respect to the
measure \equ(partition), 
%that is
%$$
%p(\xi,\h)\n_{\L,\b}(\xi)=p(\h,\xi)\n_{\L,\b}(\h),\;\;\;\;\;\; 
%\forall\xi,\h\in\O\;\; ;
%\Eq(detailed)
%$$
hence, the unique invariant measure of the process is the equilibrium
Gibbs measure.
The problem we want to study is the
way in which a system approaches the equilibrium state when it is
prepared in the configuration with all the spins
equal to minus one $(\s_0=\minusone)$ and the magnetic field $h$ is chosen
positive but small with respect to the coupling constant $J$ (${h\over J} <1$), while $\b$
is very large.
\par
In [NS1,RTMS,S1,TM1] this problem was studied for periodic
boundary conditions: it was shown that for 
$\b$ sufficiently large and $h$ small enough, depending on the size of
the lattice, the system shows metastable
behavior. This means that the system spends a long time 
$\t_{p,\b}\sim\exp (\b {4J^2\over h})$ in a phase with
negative magnetization performing random wanderings near the configuration
$\minusone$. These wanderings are characterized by the formation
of small droplets of pluses inside the sea of minuses which disappear 
quickly; their typical life time $\t_{p,\b}^s$ is much
shorter than the lifetime of the metastable state ($\t_{p,\b}^s\ll\t_{p,\b}$).
\par
A droplet of pluses will tend to grow however if it is
large enough, i. e. when its dimension is larger than 
a critical length, $l^*=[{2J\over h}]+1$, where $[a]$ is the integer
part of the real number $a$.
The exit from the metastable phase is achieved, then,  when a sufficiently 
large droplet shows up somewhere in the lattice: this 
droplet is called {\it protocritical} and
in the limit $\b\to\infty$ it is a square droplet with sides $l^*$.
When this protocritical droplet appears, 
it grows and covers the whole lattice
in a time $\t_{p,\b}^g$ which is very small compared to the 
lifetime of the metastable
phase. It has been shown that $\t_{p,\b}^s\ll\t_{p,\b}^g\ll\t_{p,\b}$.
\par
In our case  of free boundary conditions we show, rigorously in the
limit $\b\to\infty$ and via Monte Carlo simulations for $\b$ large, 
that the system exhibits metastability and that the
lifetime of this metastable phase is $\t_{\b}\sim\exp(\b{J^2\over h})$.
This exit time is 
much smaller than in the case of periodic boundary conditions.
This is a consequence of the fact that the tendency of a droplet to grow
is favoured when such a droplet has one of
its sides on the boundary of the domain or at a 
distance one from it. Indeed, we have to introduce two different critical
lengths, $\l_1$ and $\l_2$, which refer respectively to droplets close to and
far from the boundary. In the limit $\b\to\infty$ we find
$$
\l_1=\big[{J\over h}\big]+1\;\;\; {\rm and}\;\;\; 
\l_2=\big[{2J\over h}\big]+1\;\; , 
\Eq(critical-length)
$$
with $J$ and $h$ fixed such that ${J\over h} > 1$.
\par
Finally, we remark that 
the exit from the metastable phase occurs through
a critical seed which appears in one of the four corners of the
lattice. It grows in a time $\t_{\b}\sim\exp(\b(J-h))$ to cover the
whole domain. Hence, in this case the position of
the nucleation seed in the lattice can be predicted a priori. In [RKLRN] it
was observed that in the case of semiperiodic boundary conditions the
critical droplet can show up on one of the two sides where the 
periodic boundary conditions are not imposed.

\expandafter\ifx\csname sezioniseparate\endcsname\relax%
\input macro \fi
\numsec=3
\numfor=1\numtheo=1\pgn=1
\vskip 1 truecm 
\par\noindent 
{\bf 3. Numerical results.} 
\par
In this section we describe some simulation results obtained using the
Metropolis algorithm for an $M$ by $M$
square at low
temperature and small magnetic fields; the typical values which have
been used in the numerical experiments are $\b\ge 2$,  $h\le
0.5$, and $J=1$.  
This range of parameters is different from
those considered, for instance, in [RTMS,TM1] and references therein, e.g.\
in 
[RTMS] they considered the case 
$J=1$,
$\b=1.102$ and $h$ varying approximately in the range 
$0.04\le h\le 0.9$.
\par
To obtain a numerical estimate of the critical
lengths $\l_1,\l_2$ we fixed $\b$ and $h$ and 
prepared the system in the
configuration $\s=\minusone$ except for a plus droplet of size $l$
in the corner of the square;
the size of the square was chosen
$M=l+5$. We considered decreasing values of $l$ and in each
experiment we found the smallest value of $l$ such that the
droplet grew. We took this as an estimate of the critical length 
$\l_1$. The length $\l_2$ was measured in a similar way; the droplet now
being placed at the center of the lattice. 
\par
Fig. 1 presents an 
average over $60$ independent determinations of
$\l_1$ and $\l_2$ with inverse temperature $\b=10$, and 
$\b=6$ respectively.  
The solid and dashed lines represent 
the theoretical values \equ(critical-length), valid in the limit $\b\to\infty$, 
while the black circles and black squares are, respectively,  
the numerical estimates of the critical lengths $\l_1$ and $\l_2$. 
The error bars have
been omitted because the statistical errors, evaluated as the
empirical standard deviation over the square root of the number of
experiments, were found to be very small. The agreement between the
numerical result and the theoretical prediction is very good.
\par
We remark that the number of Monte Carlo steps per site (MCS) one has
to wait in order to see 
the growth of the droplet greatly increases as the value of the
magnetic field $h$ is decreased; indeed this time, for $\b$ large, is approximately given
by $\exp{\b(J-h)}$ and $\exp{\b(2J-h)}$ respectively in the case of a droplet
close to or far from the boudary of the domain.
The smallest magnetic field we have
considered is $h=0.02$: in this case we have set $M=56$ and the number
of MCS needed to see the shrinking or growth of the droplet was
approximatively $10^5$.
\par
In Fig. 2 we have
plotted the magnetization per spin of the whole box $m_0$ and
the magnetization per spin $m_1,\; m_2\; m_3$ and $m_4$ evaluated in
four square boxes of side $\l_1$ placed at the four corners of
the lattice, 
as functions of the number of iterations (time) in a single
history of the system obtained after preparing the system in the starting
configuration $\minusone$. In both pictures the solid line
represents the magnetization of the whole lattice $m_0$, while the
other lines refer to $m_1,\; m_2\; m_3$ and $m_4$. The top graph 
in Fig. 2 refers to the case where $\b=3$,
$h=0.24$ ($\l_1=5$) and $M=16$; the one below has been obtained
with $\b=2$, $h=0.14$ ($\l_1=8$) and $M=32$. 
\par
In both cases it is clear that the system stays for a long time
(about $10^4$ MCS) in the configuration
$\minusone$: the thermal fluctuations, visible for $\b=2$,
are negligible for $\b=3$.  After this long
time the magnetization in one of the corner boxes flips to one
(nucleation of the protocritical droplet); once this rare event has
happened, all the other magnetizations start to grow and quickly reach
the value $+1$; that is the system quickly reaches the equilibrium state.
\par
Finally, we have evaluated the lifetime of the metastable state. In
[RTMS,MT] and references therein the lifetime of the metastable state
has been estimated at temperature $T=0.8 T_c$, where $T_c$ is the
Onsager critical temperature, corresponding to our $\b=1.102$. As the
magnetic field is varied, four different regimes are detected (see [RTMS]
for a detailed description of these different regimes); at low values of
the magnetic field they find that the logarithm of the lifetime of the
metastable state is a linear function of the inverse of the magnetic field.
They call this the ``single-droplet region'', meaning that the nucleation of
the stable state is achieved via the formation of a single cricical
droplet, see also [RKLRN,TM2]

	In Fig. 3 we have plotted our numerical measurements of the escape
time $\tau_\beta$ versus the inverse of the magnetic field.  It is clear
that when 
the temperature is decreased the numerical results approach the
theoretical value
$$
{1\over\b}\log\t_{\b}\approx {1\over h}\;\; 
\Eq(approx-escape)
$$
which by Theorem 1.1 is valid in the limit of zero
temperature. 
The numerical data have been fitted with a linear
function ${1\over\b}\log\t_{\b}=m_{\b} {1\over h} + n_{\b}$; the
values of $m_{\b}$ and $n_{\b}$ 
are listed in Table 1.  It is clear that the trend
with increasing $\b$ is correct, although the value \equ(approx-escape) 
would be reached only for much larger values of $\b$. 
Such simulation would require very long computer
time.
\par
The results in Fig. 3 have been obtained in the case $M=32$ and are the
average of $60$ different histories.
We have checked that very similar results are obtained if one considers
larger domains (for instance $M=64, 128$) but have not performed
extensive statistics in these situations because the
behavior of $\t_{\b}$ at small values of the magnetic field and very
low temperatures does not depend on the size of the lattice (this is
confirmed by the results in [RTMS], see Fig. 2 there).
Finally, we remark that the linearity of the logarithm of $\t_{\b}$ is
lost when $1/h$ is small enough, this is because for $h$ large enough
the system is in the multi-droplet regime. 

\expandafter\ifx\csname sezioniseparate\endcsname\relax%
\input macro \fi
\numsec=4
\numfor=1\numtheo=1\pgn=1
\vskip 1 truecm
\par\noindent
{\bf 4. Rigorous results.}
\par
We list some definitions and notations which are necessary to discuss
our rigorous results:
\par\noindent
$1.$   The spin configurations $\s,\h\in\O$,  are called  
{\it nearest neighbor} configurations
iff $\exists x\in\L$ such that $\h=\s^x$, where $\s^x$ is the configuration
obtained by flipping in $\s$ the spin at site $x$. 
A {\it path} is a sequence of configurations
$\s_0\s_1...\s_n$ such that for $i=1,...,n-1,\;\; \s_{i-1}$ and
$\s_i$ are nearest neighbors. A path $\s_0\s_1...\s_n$ is called
{\it downhill} iff $H(\s_{i+1})\le H(\s_i)$ for $i=0,1,...n-1$. 
\par\noindent
$2.$ Given $A\subset\O$, $\h\in\O$ we define the
{\it hitting time}
$$
\t_A^{\h}=\inf\{t\ge 0:\;\s_t^{\h}\in A\}\;\; .
\Eq(hitting)
$$
\par\noindent
$3.$ A {\it local minimum} of the Hamiltonian is a
configuration $\s$ such that one has 
$$
H(\s^x)>H(\s)\;\;\;\;\;\;\;\;\forall x\in\L\;\; .
\Eq(local-minimum)
$$
A local minimum will also be called a 
{\it stable configuration} because
starting from it the system will not move for a time which is exponentially
long in the inverse temperature $\b$. 
\par\noindent
$4.$ The set of all the local minima is denoted by $\cM\subset\O$.
\par\noindent
$5.$ Given $\h\in\cM$, we consider the process starting from
$\h$ and say
$$
\eqalign{
\h\; subcritical\;&\Longleftrightarrow\lim_{\b\to\infty}
P(\t_{\minusone}^{\h} < \t_{\plusone}^{\h})=1\cr
\h\; supercritical\;&\Longleftrightarrow\lim_{\b\to\infty}
P(\t_{\plusone}^{\h} <\t_{\minusone}^{\h})=1\cr}
\Eq(criticality)
$$
where $P(X)$ is the probability of the event $X$.
\par\noindent
$6.$ Given $\s\in\cM$ we define its {\it basin of attraction}
$$
B(\s)=\{\h\in\O:\; {\rm all\; downhill\; paths\; starting\;
from\;}\h\; {\rm end\; in\;}\s\}\;\; .
\Eq(basin)
$$
We observe that a downhill path $\o=\s_0\s_1...\s_n$ necessarily ends in a
local minimum if
$$
n\ge\cT={\max_{\h\in\O}H(\h)-\min_{\h\in\O}H(\h)\over h}\;\; .
$$
\par\noindent
$7.$ Given $\cG\subset\O$, $\cG$ is {\it connected} iff
$\forall\h,\s\in\cG$ $\exists$ a path $\o\subset\cG$ starting from
$\s$ and ending in $\h$; we will say that this path {\it
connects} $\s$ {\it to} $\h$.
\par\noindent
$8.$ Given a connected set $\cG\subset\O$, we will call
the {\it boundary of} $\cG$ the set
$$
\partial\cG=\{\h\in\O:\;\h\not\in\cG,\;\exists
x\in\L:\;\h^x\in\cG\}
\;\;.
\Eq(bound)
$$
\par\noindent
$9.$ Given two rectangles $R_1$ and $R_2$ on the dual lattice
$\L+({1\over 2},{1\over 2})$, we say that $R_1$ and $R_2$ are {\it
interacting} rectangles iff 
$R_1$ and $R_2$ intersect or are separated by one lattice spacing.
If two such rectangles have only two corners at
distance one, then they are considered not interacting.
\par
Let $h>0$ and start the
system from the 
configuration $\s_0=\minusone$. We will now describe how the stable
configuration $\plusone$ is approached and we will evaluate how long the
system remains in the metastable phase. To do this we
first characterize the local minima of the hamiltonian
\equ(hamiltonian).
\vskip 0.5 truecm
\par\noindent
{\bf Lemma 1.1}
\par\noindent
Let us consider model \equ(hamiltonian) with $J>h>0$ and $M > 2$.  Then 
$\s\in\cM$ iff $\s(x)=-1\;\forall x\in\L$ except
for sites which are inside some 
rectangles $R_1,...,R_n$ laying on the dual lattice 
$\L+({1\over 2},{1\over 2})$ such that $\forall i,j=1,...,n$ and $i\not= j$
\item{$i)$} $R_i$ and $R_j$ are not interacting 
\item{$ii)$} $R_i$ has sides longer than two
\item{$iii)$} $R_i$ cannot have one of its sides on the ``border''
of $\L$ and one of its two other sides perpendicular to this one at
a distance one from the border (see Fig. 4).
\vskip 0.5 truecm
\par\noindent
{\bf Remarks.} 
\par\noindent
$1.$ We denote by $\cR(l_1,l_2)$ with $2\le l_1,l_2\le M$ the set of all
configurations with all spins $-1$ except
for those inside a rectangle with sides $l_1$ and $l_2$ and such that
they are local minima.
\par\noindent
$2.$ Given a local minimum $\s\in\cR(l_1,l_2)$ we denote it by 
$R_{l,m}$ where
$$
l=\min\{l_1,l_2\}\;\;\;\;{\rm and}\;\;\;\;
m=\max\{l_1,l_2\}\;\; .
\Eq(sides)
$$
\par\noindent
$3.$ We denote by $\cR\subset\cM$ the set of all local minima
containing only one rectangle of pluses.
\par
Now, we state under which conditions a local minimum is subcritical,
that is we describe the evolution of the system starting from a stable
configuration. Consider a local minimum $R_{l,m}\in\cR$: 
\vskip 0.5 truecm
\par\noindent
{\bf Lemma 1.2}
\par\noindent
When each side of $R_{l,m}$ is at least at distance two from the border of
the lattice, then given $\e>0$ one has
\item{$i)$} $l<\l_2\;\Longrightarrow\; R_{l,m}$ is subcritical
and
$$
P(e^{\b(l-1)h-\b\e}<\t_{\minusone}^{R_{l,m}}<e^{\b(l-1)h+\b\e})\;\;\;
{\buildrel \b\to\infty\over\longrightarrow}\;\;\; 1
\Eq(sub-time-1)
$$
\item{$ii)$} $l\ge \l_2\;\Longrightarrow\; R_{l,m}$ is supercritical and
$$
P(e^{\b(2J-h)-\b\e}<\t_{\plusone}^{R_{l,m}}<e^{\b(2J-h)+\b\e})\;\;\;
{\buildrel \b\to\infty\over\longrightarrow}\;\;\; 1
\Eq(super-time-1)
$$
\vskip 0.3 truecm
\par\noindent
When at least one of the sides of $R_{l,m}$ is at distance one from the 
border of the square or is laying on it, then given $\e>0$ one has
\item{$i)$} $l<\l_1,\; m\le M-1\;\Longrightarrow\; R_{l,m}$ is subcritical
and
$$
P(e^{\b(l-1)h-\b\e}<\t_{\minusone}^{R_{l,m}}<e^{\b(l-1)h+\b\e})\;\;\;
{\buildrel \b\to\infty\over\longrightarrow}\;\;\; 1
\Eq(sub-time-2)
$$
\item{$ii)$} $l\ge \l_1\;\Longrightarrow\; R_{l,m}$ is supercritical and
$$
P(e^{\b(J-h)-\b\e}<\t_{\plusone}^{R_{l,m}}<e^{\b(J-h)+\b\e})\;\;\;
{\buildrel \b\to\infty\over\longrightarrow}\;\;\; 1
\Eq(super-time-2)
$$
\vskip 0.5 truecm
\par\noindent
{\bf Remarks.} 
\par\noindent
$1.$ In the second case, if $m=M$ then the local minimum is
supercritical, no matter how long the smallest side $l$ is. 
\par\noindent
$2.$ If one considers a local minimum with more than one
rectangle, then this local minimum is subcritical iff all its
rectangles are ``subcritical'' (abuse of language).
\par\noindent
$3.$ It is possible to prove a stronger version  of Lemma 1.2 that
gives rise to a more detailed description of the contraction or the
growth of the droplet (see Lemma 3 and 4 in [KO1], Theorem 1 in [NS1]).
\vskip 0.5 truecm
\par
The proof of this lemma is the standard one, see, for instance,
[NS1,KO1,KO2] or the proof of Proposition 4.1 in [CO]. One has to
consider the basin of attraction of the local minimum $R_{l,m}$ and
work out the minimum of the energy on its boundary. Once this has been
done, everything follows via the general arguments in Proposition 3.7 in [OS1].
The difference with respect to the case of periodic boundary
conditions is that one must take into account
situations like the one depicted in Fig. 5, that are absent in the
Ising model with periodic boundary conditions.
\par
We can formulate, now, the theorem which describes the exit from the
metastable phase: this theorem states that with high probability the
system will visit a particular configuration $\cP$ before reaching
$\plusone$ and that the exit time is dominated by the time the system
needs to reach $\cP$.
\par
This {\it protocritical} configuration $\cP$ is such that all the spin
are minuses except those in the union of  a $\l_1\times (\l_1-1)$
rectangle, one  of whose corners coincides with one of the corner of the
domain, and a unit square laying on the border of $\Lambda$ and touching
the longer side of the rectangle (see Fig. 6).  
\par
Setting 
$$
\G=H(\cP)-H(\minusone)=J+2J\l_1-h(\l_1^2-\l_1+1)\;\; ;
\Eq(prot-energy)
$$
we consider the process $\s_t$ starting from $\minusone$ and define
$\tau_{-1}$ as the last time the configuration was $\minusone$ before it
became $\plusone$, 
$$
{\bar\t}_{\minusone}=\sup\{t<\t_{\plusone}:\; \s_t=\minusone\}.
\Eq(last-visit)
$$
Similarly we call ${\bar\t}$ the first time after ${\bar\t_{\minusone}}$
when  $\s_t=\cP$, 
$$
{\bar\t}_{\cP}=\inf\{t>{\bar\t}_{\minusone}:\; \s_t=\cP\}\;\; .
\Eq(prot-visit)
$$
Finally, we can state the following theorem:
\vskip 0.5 truecm
\par\noindent
{\bf Theorem 1.1}
\par\noindent
Given $\e>0$,
\item{$i)$} $P_{\minusone}({\bar\t}_{\cP}<\t_{\plusone})\;\;\;
{\buildrel \b\to\infty\over\longrightarrow}\;\;\; 1$ 
\item{$ii)$} 
$P_{\minusone}(e^{\b\G-\b\e}<\t_{\plusone}<e^{\b\G+\b\e})\;\;\;
{\buildrel \b\to\infty\over\longrightarrow}\;\;\; 1$ 
\vskip 0.5 truecm
\par
{}From Lemma 1.2 and Theorem 1.1 we have a rather accurate description
of the system in the metastable phase: starting from $\minusone$ the
system will spend a lot of time ``close'' to this configuration; sometimes
small droplet of pluses appear, but the system quickly goes back to
$\minusone$. Only after a long time, compared to the time these
fluctuations need, the system will nucleate the protocritical droplet
$\cP$ and it will then reach the stable phase in a relatively short time.
\par
It is possible to give a more detailed description of the first
excursion from $\minusone$ to $\plusone$; one could state a result
similar to the one in [S1] (see also Theorem 3 in [KO1]). We do not
enter in the details of this construction for the case of free
boundary conditions, because no relevant difference appears with
respect to the case of periodic boundary conditions. Roughly speaking
one can say that that the system, during the excursion from
$\minusone$ to $\plusone$, will follow a rather
well-defined sequence of configurations made up of growing rectangular,
almost square, droplets located at one of the four corners of $\Lambda$.
\par
The proof of Theorem 1.1 is now sketched: as in [KO1] we can define a
set $\cA$ satisfying some properties which will be listed below. 
The construction of the set $\cA$ is exactly as in
[KO1], except for the fact that whenever there is  a rectangle with one side
on the ``border''  and another side 
 at distance one from the border of $\L$, the rectangle is enlarged so that
the latter side also touches 
 the border (see e.g.\ Fig. 4).
\par
The relevant properties of the set $\cA$ are the following:
\vskip 0.5 truecm
\par
\item{$i)$} $\cA$ is connected; $\minusone\in\cA$ and
$\plusone\not\in\cA$.
\item{$ii)$} There exists a path $\o$ connecting $\minusone$ with
$\cP$ contained in $\cA$ such that
$$
H(\s)<H(\cP)\;\;\forall\s\in\o,\; \s\not=\cP\;\; .
$$
There exists a path $\o'$ connecting $\cP$ with
$\plusone$ contained in $\cA^c$ such that
$$
H(\s)<H(\cP)\;\;\forall\s\in\o',\; \s\not=\cP\;\; .
$$
\item{$iii)$} The minimal energy in $\partial\cA$ is attained for the
``protocritical'' configuration; namely
$$
\min_{\s\in\cA} (H(\s)-H(\minusone))=H(\cP)-H(\minusone)=\G
\Eq(min-bound)
$$
and
$$
\min_{\s\in\cA\setminus\{\cP\}} (H(\s)-H(\cP))>0\;\; .
\Eq(only-min-bound)
$$
\item{$iv)$} With probability greater than zero, uniformly in $\b$,
the system starting from $\cP$ will reach $\plusone$ before visiting
$\minusone$; namely, given $\e>0$
$$
P(\t_{\plusone}^{\cP}<\t_{\minusone}^{\cP})\ge e^{-\e\b}
$$
and
$$
P(\t_{\plusone}^{\cP}<e^{\b(J-h)+\b\e}|
\t_{\plusone}^{\cP}<\t_{\minusone}^{\cP})
{\buildrel \b\to\infty\over\longrightarrow} 1 \;\; .
$$
\vskip 0.5 truecm
\par\noindent
Using properties $i)-iv)$ and Propositions 3.4, 3.7 in [OS1] we
get Theorem 1.1.

\expandafter\ifx\csname sezioniseparate\endcsname\relax%
\input macro \fi
\numsec=5
\numfor=1\numtheo=1\pgn=1
\vskip 1 truecm
\par\noindent
{\bf 5. Concluding Remarks.}
\par
In this paper we have studied metastability in the
 two-dimensional Ising model on an $M$ by $M$ square with free boundary
conditions rigorously in the limit $\beta \to \infty$ and via 
Monte Carlo simulations at finite temperatures.  We found good agreement
between the theoretical predictions and the simulations and for a large
range of $h$ and low enough temperatures.  The qualitative agreement
persists even above one half of $T_c$.  
\par
Compared to 
periodic boundary conditions there are two relevant differences: $i)$
the critical length of the droplet and hence the life time of the
metastable phase is much shorter;
$ii)$ the protocritical droplet is always at 
 one of the four corners
of the square.  
\par
It is clear that our analysis applies equally well to a rectangular domain
with sufficiently long sides. In fact the basic approach carries over, in
principle, to a general domain with general boundary conditions.  The
protocritical domain will always form, when $\beta \to \infty$ at the place
(or places) where the energy cost, $H(\cP) - H(\sigma_{\minusone})$, is
minimal.

\vskip 1 truecm
\par\noindent
{\bf Acknowledgements.}
\par
We are indebted to J. Marro for pointing out to us the interest of this
case in modeling relaxation in some magnetic particles [M], and to Per Arne
Rikvold for useful comments.  
One of the authors (E.C.) wishes to express his thanks  
to the Mathematics Department of Rutgers
University for its very kind hospitality and to Enzo Olivieri
for useful discussions. E.C. also thanks also Istituto Nazionale di Fisica 
Nucleare - Sezione di Bari and Dipartimento di Fisica dell'Universit\`a 
degli Studi di Bari for their financial support. Work at Rutgers
was supported by NSF Grant DMR 95 - 23266.

%Bibliografia
\vfill\eject
\vskip 2 truecm
\centerline {\bf References.}
\vskip 0.4 truecm
\par\noindent

\item{[B]} K. Binder, Phys. Rev. B {\bf 8}, 3423 (1973).

\item{[BC]} G. Ben Arous, R. Cerf, ``Metastability of the three
dimensional Ising model on a torus at very low temperatures.''
Electronic Journal of Probability {\bf 1}, 1-55 (1996).

\item{[BM]} K. Binder, H. M\"uller-Krumbhaar, Phys. Rev. B {\bf 9},
2328 (1974).

\item{[BS]} K. Binder, E. Stoll, Phys. Rev. Lett. {\bf 31}, 47 (1973).

\item{[CGOV]} M. Cassandro, A. Galves, E. Olivieri, M.E. Vares,
``Metastable behavior of stochastic dynamics: A pathwise approach.''
Journ. Stat. Phys. {\bf 35}, 603-634 (1984). 

\item{[CO]} E.N.M. Cirillo, E. Olivieri,
``Metastability and nucleation for the Blume-Capel model. Different mechanisms 
of transition." Journ. Stat. Phys. {\bf 83}, 473-554 (1996).

\item{[FGRN]} T. Fiig, B.M. Gorman, P.A. Rikvold, M.A. Novotny,
Phys. Rev. E {\bf 50}, 1930 (1994).

\item{[GRN]} C.C.A. G\"unther, P.A. Rikvold, M.A. Novotny,
Phys. Rev. Lett. {\bf 71}, 3898 (1993); Physica A {\bf 212}, 194--229 (1994).

\item{[I]} S.N. Isakov, ``Nonanalytic feature of the first order phase 
transition in the Ising model." Comm. Math. Phys. {\bf 95}, 427-443 (1984).

\item{[KO1]} R. Kotecky, E. Olivieri, ``Droplet dynamics for asymmetric
Ising model.'' 
Journ. Stat. Phys. {\bf 70}, 1121-1148 (1993).

\item{[KO2]} R. Kotecky, E. Olivieri, ``Shapes of growing droplets -
a model of escape from a metastable phase.'' 
Journ. Stat. Phys. {\bf 75}, 409-507 (1994).

\item{[Li]} T.M. Ligget, ``Interacting Particle System",
(Springer--Verlag, New York).

\item{[LR]} O. Lanford, D. Ruelle, 
``Observable at infinity and states with short range correlations in 
statistical mechanics." Commun. Math. Phys. {\bf 13}, 194-215 (1969).

\item{[M]} J. Marro, J.A. Vacas, ``Discontinuous particle demagnetization
at low temperature." Preprint.

\item{[MOS]}  F. Martinelli, E. Olivieri, E. Scoppola,
``Metastability and exponential approach to equilibrium for
low temperature stochastic Ising models.'' Journ. Stat. Phys. {\bf 61}, 
1105 (1990).

\item{[NO]} F.R. Nardi, E. Olivieri, ``Low temperature Stochastic Dynamics 
for an Ising Model with Alternating Field." Markov Proc. and Rel. Fields
{\bf 2}, 117-166 (1996).

\item{[NS1]}  E.J. Neves, R.H. Schonmann,
``Critical Droplets and Metastability for a Glauber Dynamics
at Very Low Temperatures."  
Comm. Math. Phys. {\bf 137}, 209 (1991).

\item{[NS2]}  E.J. Neves, R.H. Schonmann, ``Behavior of droplets for a 
class of Glauber dynamics at very low temperatures.'' 
Prob. Theor. Rel. Fields {\bf 91}, 331 (1992).

\item{[OS1]} E. Olivieri, E. Scoppola, ``Markov chains with exponentially 
small transition probabilities: First exit problem from a general domain - I. 
The reversible case.'' 
Journ. Stat. Phys. {\bf 79}, 613-647 (1995).

\item{[OS2]} E. Olivieri, E. Scoppola, ``Markov chains with exponentially 
small transition probabilities: First exit problem from a general domain - II. 
The general case.'' Journ. Stat. Phys. {\bf 84}, 987-1041 (1996).

\item{[PL]} O. Penrose, J.L. Lebowitz, ``Towards a rigorous
molecular theory of metastability." In Fluctuation Phenomena (second edition). 
E. W. Montroll, J. L. Lebowitz, editors. North-Holland Physics Publishing, 
(1987);  O. Penrose, J.L. Lebowitz, ``Molecular theory of metastability: 
An update." Appendix to the reprinted edition of the article ``Towards a 
rigorous molecular theory of metastability" by the same authors. In: 
Fluctuation Phenomena (second edition). E.W. Montroll, J.L. Lebowitz,
(eds.) Amsterdam: North--Holland Physics Publishing 1987.

\item{[PS1]} V. Privman, L.S. Schulman, J. Phys. A {\bf 15}, L321
(1982).

\item{[PS2]} V. Privman, L.S. Schulman, Journ. Stat. Phys. {\bf 31},
205 (1982).

\item{[RKLRN]} H.L. Richards, M. Kolesik, P.A. Lindg\aa rd, P.A. Rikvold,
M.A. Novotny, ``Effects of boundary conditions on magnetization switching in
kinetic Ising moldels of nanoscale ferromagnets." 
Phys. Rev. B {\bf 55}, 11521 (1997).

\item{[RSNR]} H.L. Richards, S.W. Sides, M.A. Novotny, P.A. Rikvold,
Journ. Magnetism Magn. Materials {\bf 150}, 37-50 (1995).

\item{[RTMS]} P.A. Rikvold, H. Tomita, S. Miyashita, S.W. Sides,
Phys. Rev. E {\bf 49}, 5080 (1994).

\item{[S1]} R.H. Schonmann, 
``The pattern of escape from metastability of a stochastic Ising model.'' 
Comm.  Math.  Phys. {\bf 147}, 231-240 (1992).

\item{[S2]} R.H. Schonmann,
``Slow droplet-driven relaxation of stochastic Ising
models in the vicinity of the phase coexistence region." Comm. Math. Phys. 
{\bf 161}, 1-49 (1994).

\item{[SS]} S. Shlosman, R.H. Schonmann, preprint UCLA (1994).

\item{[TM1]} H. Tomita, S. Miyashita, Phys. Rev. B {\bf 46}, 8886
(1992).

\item{[TM2]} H. Tomita, S. Miyashita, ``Statistical properties of the 
relaxation processes of metastable states in the kinetics Ising model (II) -
Free boundary conditions." Preprint.

\vfill\eject

%Tabella 1
\centerline{\bf Table 1}
\vskip 2 truecm
\centerline{     % Questo hfil serve a centrare la tabella
%\hfil{     % Questo hfil serve a centrare la tabella
\vbox{\offinterlineskip % Offinterlineskip annulla il lineskip
\hrule     %traccia una riga orizzontale
{\phantom .}
\hrule     %traccia una riga orizzontale
\halign{&\vrule#&
   \strut\quad\hfil#\quad\cr
height2pt
&\omit&&\omit&&\omit&\cr
&\hfil $\b$\hfil&
&\hfil $m_{\b}$\hfil&
&\hfil $n_{\b}$\hfil&\cr
height2pt&\omit&&\omit&&\omit&\cr
\noalign{\hrule}
\noalign{
{\phantom .}
\hrule}     %traccia una riga orizzontale
height2pt&\omit&\cr
&1.00&& 0.216 && 1.816 &\cr
&1.25&& 0.312 && 1.173 &\cr
&1.50&& 0.483 && 0.669 &\cr
&1.75&& 0.629 && 0.378 &\cr
&2.00&& 0.782 && 0.083 &\cr
height2pt&\omit&&\omit&&\omit &\cr}
\hrule
{\phantom .}
\hrule     %traccia una riga orizzontale
}}
\vskip 5 truecm
\noindent
\centerline{
\vbox{
\hsize=13truecm
\baselineskip 0.35cm
\noindent
{\ninebf Table 1:} {\ninerm \quad Results of the linear fit 
${1\over\b}\log\t_{\b}=m_{\b} {1\over h} + n_{\b}$ of the data in Fig. 3.
For $\beta \to \infty$, $m_\beta \to 1$, $n_\beta \to \infty$.}}}

%Inserimento Fig.1
\vfill\eject
\vskip 0.5 truecm
\eqfig{600pt}{570pt}{
\ins{150pt}{140pt}{\vbox
{
\hsize=13truecm
\baselineskip 0.35cm
\noindent
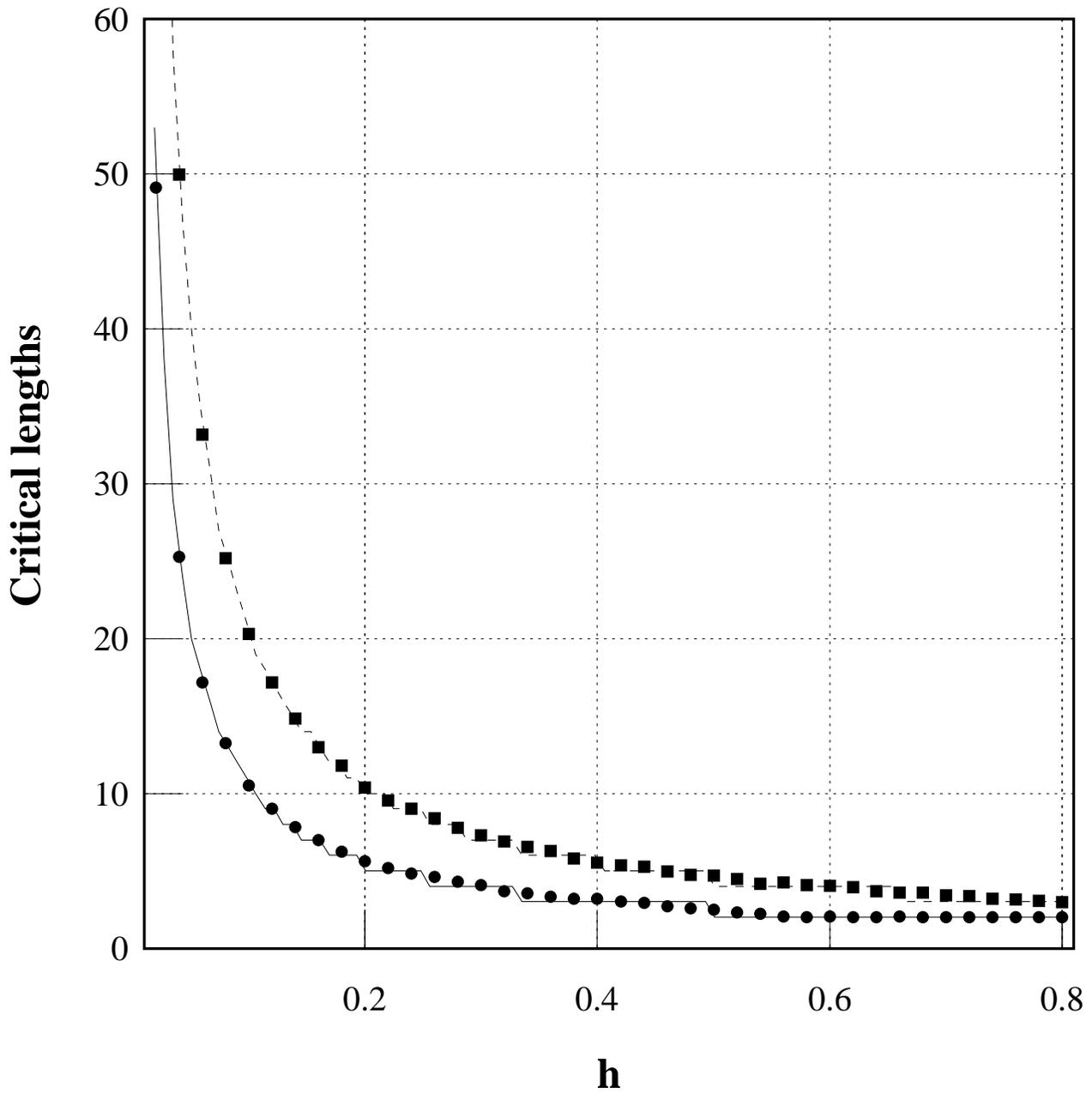
{\ninebf Fig. 1:}{\ninerm \quad The critical lengths $\l_1$ and $\l_2$ 
as functions of the magnetic field $h$. The solid and dashed lines represent 
the theoretical prediction \equ(critical-length) in the limit
$\b\to\infty$; the black circles and the black squares are, respectively, 
the numerical estimates of $\l_1$ at $\b=10$ and $\l_2$
at $\b=6$.}
}}
}
{isingfree.fig1}{}
\vfill\eject
%%%%%%%%%Fine Fig.1

%Inserimento Fig.2
\vskip 0.5 truecm
\eqfig{600pt}{570pt}{
\ins{150pt}{140pt}{\vbox
{
\hsize=13truecm
\baselineskip 0.35cm
\noindent
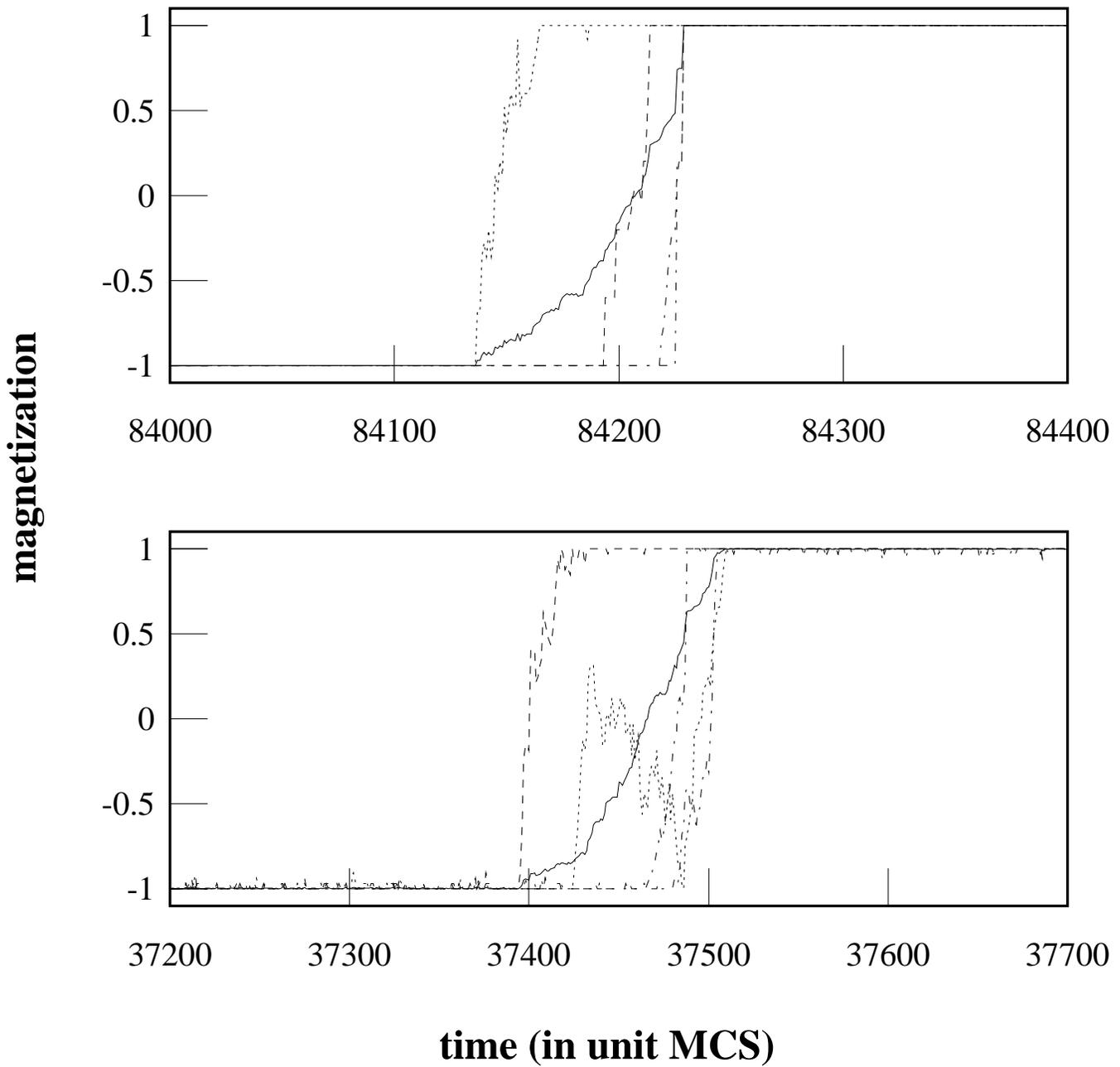
{\ninebf Fig. 2:}{\ninerm \quad The solid line represents the average 
magnetization of the whole lattice $m_0$; the other lines represent 
$m_1,\; m_2\; m_3$ and $m_4$. The top figure corresponds 
to the case $\b=3$,
$h=0.24$ ($\l_1=5$) and $M=16$; the bottom one 
to $\b=2$, $h=0.14$ ($\l_1=8$) and $M=32$.}
}
}}
{isingfree.fig2}{}
\vfill\eject
%%%%%%%%%Fine Fig.2

%Inserimento Fig.3
\vskip 0.5 truecm
\eqfig{600pt}{570pt}{
\ins{150pt}{140pt}{\vbox
{
\hsize=13truecm
\baselineskip 0.35cm
\noindent
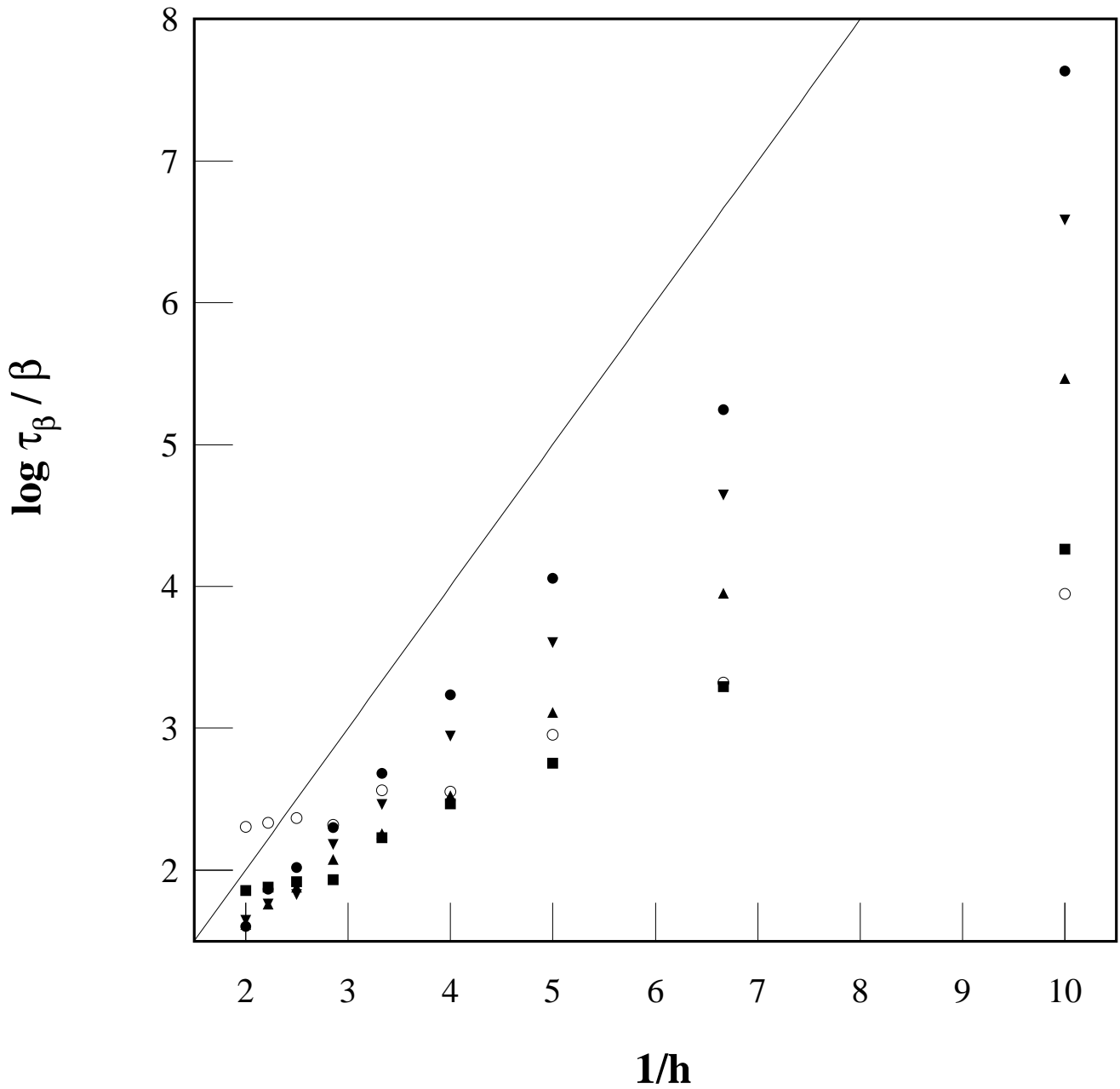
{\ninebf Fig. 3:}{\ninerm \quad Numerical estimates of
$\log\t_{\b}/\b$ plotted versus $1/h$ for
different values of the inverse temperature $\b$. Empty circles, black squares,
black upward triangles, black downward triangles and black circles
refer respectively to $\b=1.00,1.25,1.50,1.75,2.00$. All the results
are averages over $60$ different histories for $M=32$. The solid line is
the graph of a linear function with slope equal to 1.}
}
}}
{isingfree.fig3}{}
\vfill\eject
%%%%%%%%%Fine Fig.3

%Inserimento Fig.4
\vskip 0.5 truecm
\eqfig{300pt}{100pt}{
\ins{50pt}{35pt}{$+$}
\ins{250pt}{35pt}{$+$}
\ins{50pt}{65pt}{$-$}
\ins{250pt}{65pt}{$-$}
\ins{130pt}{45pt}{$\buildrel {\scriptstyle\D H=-h}\over\longrightarrow$}
}
{isingfree.fig4}{}
\vskip 0.5 truecm
\centerline{
\vbox
{
\hsize=13truecm
\baselineskip 0.35cm
\noindent
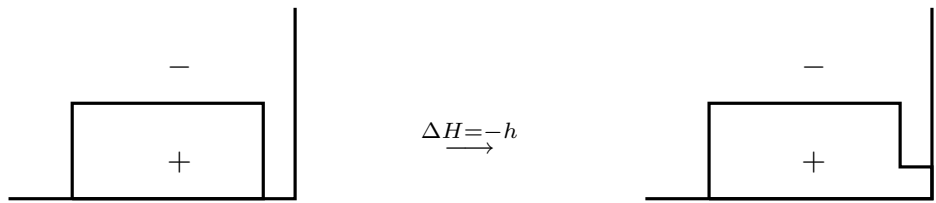
\centerline{{\ninebf Fig. 4:}{\ninerm \quad Example of a rectangular
droplet which is not a local minimum.}}
}}
\vskip 0.5 truecm
\vfill\eject
%%%%%%%%%Fine Fig.4

%Inserimento Fig.5
\vskip 0.5 truecm
\eqfig{250pt}{250pt}{
\ins{50pt}{35pt}{$+$}
\ins{250pt}{35pt}{$+$}
\ins{95pt}{50pt}{$-$}
\ins{295pt}{50pt}{$-$}
\ins{145pt}{45pt}{$\buildrel {\scriptstyle\D H=J-h}\over\longrightarrow$}
\ins{70pt}{190pt}{$+$}
\ins{270pt}{190pt}{$+$}
\ins{45pt}{160pt}{$-$}
\ins{250pt}{160pt}{$-$}
\ins{140pt}{175pt}{$\buildrel {\scriptstyle\D H=J-h}\over\longrightarrow$}
}
{isingfree.fig5}{}
\vskip 0.5 truecm
\centerline{
\vbox
{
\hsize=13truecm
\baselineskip 0.35cm
\noindent
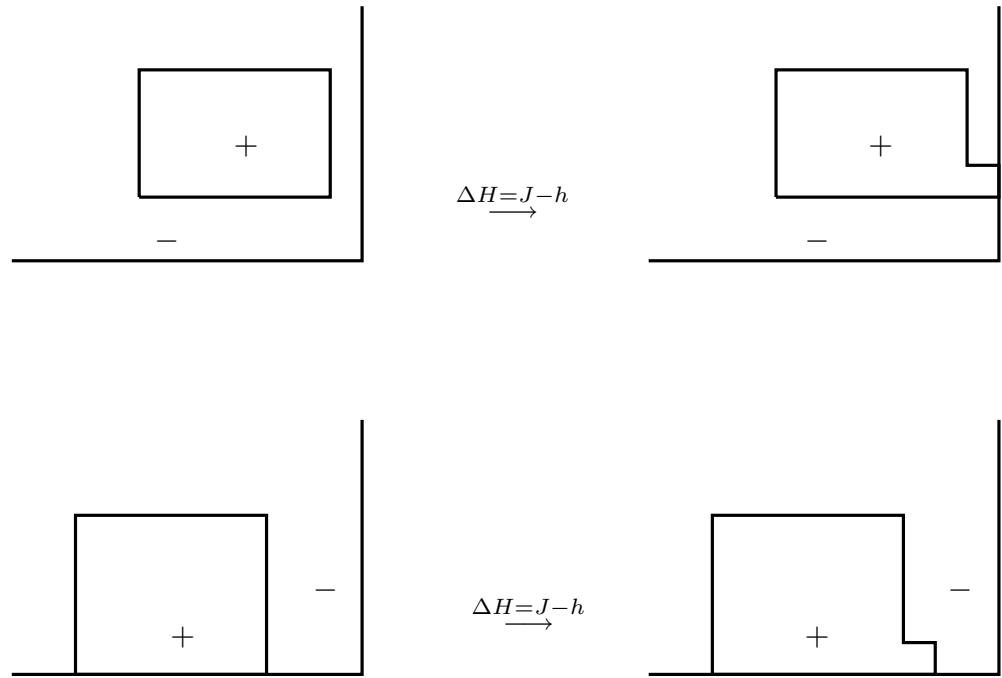
\centerline{{\ninebf Fig. 5:}{\ninerm \quad Mechanisms of growth
peculiar to the Ising model with free boundary conditions.}}
}}
\vskip 0.5 truecm
\vfill\eject
%%%%%%%%%Fine Fig. 5

%Inserimento Fig.6
\vskip 0.5 truecm
\eqfig{130pt}{130pt}{
\ins{40pt}{40pt}{$+$}
\ins{70pt}{80pt}{$-$}
\ins{130pt}{70pt}{$\L$}
\ins{65pt}{35pt}{$\l_1-1$}
\ins{40pt}{15pt}{$\l_1$}
}
{isingfree.fig6}{}
\vskip 0.5 truecm
\centerline{
\vbox
{
\hsize=13truecm
\baselineskip 0.35cm
\noindent
\centerline{{\ninebf Fig. 6:}{\ninerm \quad Protocritical droplet $\cP$.}}
}}
\vskip 0.5 truecm
\vfill\eject
%%%%%%%%%Fine Fig.6

\vfill\eject
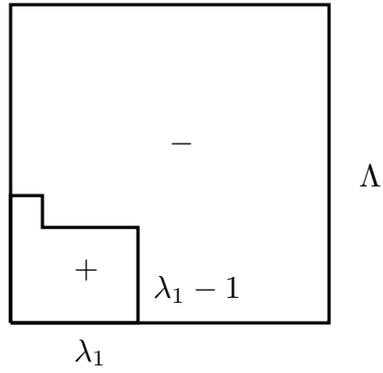
\bye